\documentclass[epj]{svjour}
\usepackage{graphics}
\begin{document}
\title{Non-uniform doping across the Fermi surface of NbS$_2$ intercalates}
\author{C. Battaglia\inst{1} \and H. Cercellier\inst{1} \and L. Despont\inst{1} \and C. Monney\inst{1} \and M. Prester\inst{2} \and H. Berger\inst{3} \and L. Forr\`{o}\inst{3} \and M.G. Garnier\inst{1} \and P. Aebi\inst{1}}
\institute{Institut de physique, Universit\'{e} de Neuch\^{a}tel,
Switzerland \and Institute of physics, Zagreb, Croatia \and
Institute of physics of complex matter, Ecole polytechnique
f\'{e}d\'{e}rale de Lausanne, Switzerland}

\date{Received: date / Revised version: date}
%
\abstract{Magnetic ordering of the first row transition metal
intercalates of NbS$_2$ due to coupling between the conduction
electrons and the intercalated ions has been explained in terms of
Fermi surface nesting. We use angle-resolved photoelectron
spectroscopy to investigate the Fermi surface topology and the
valence band structure of the quasi-two-dimensional layer compounds
Mn$_{1/3}$NbS$_2$ and Ni$_{1/3}$NbS$_2$. Charge transfer from the
intercalant species to the host layer leads to non-uniform, pocket
selective doping of the Fermi surface. The implication of our
results on the nesting properties are discussed.
} 
\maketitle
\section{Introduction}
\label{intro}

\begin{figure}
\includegraphics{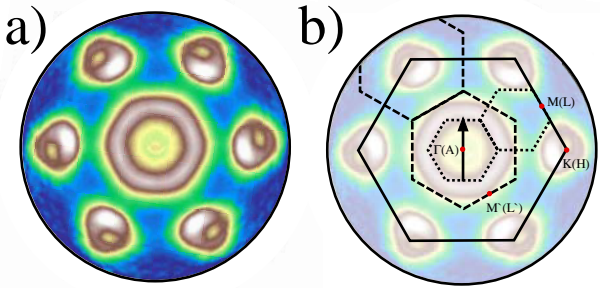}
\caption{a) Fermi surface map of Mn$_{1/3}$NbS$_2$ at room
temperature. The corresponding color scale is given in Fig.
\ref{fig:2}a). b) A sketch of the Brillouin zone of the host
compound (full line) and of the $\sqrt{3}\times\sqrt{3}$ (dashed
line) and $3\times3$ (dotted line) supercells with high symmetry
points. High symmetry points in parenthesis are located on the top
face of the hexagonal bulk Brillouin zone. The nesting vector
corresponding approximately to the $3\times3$ superlattice is also
indicated.} \label{fig:1}
\end{figure}

\begin{figure*}
\includegraphics{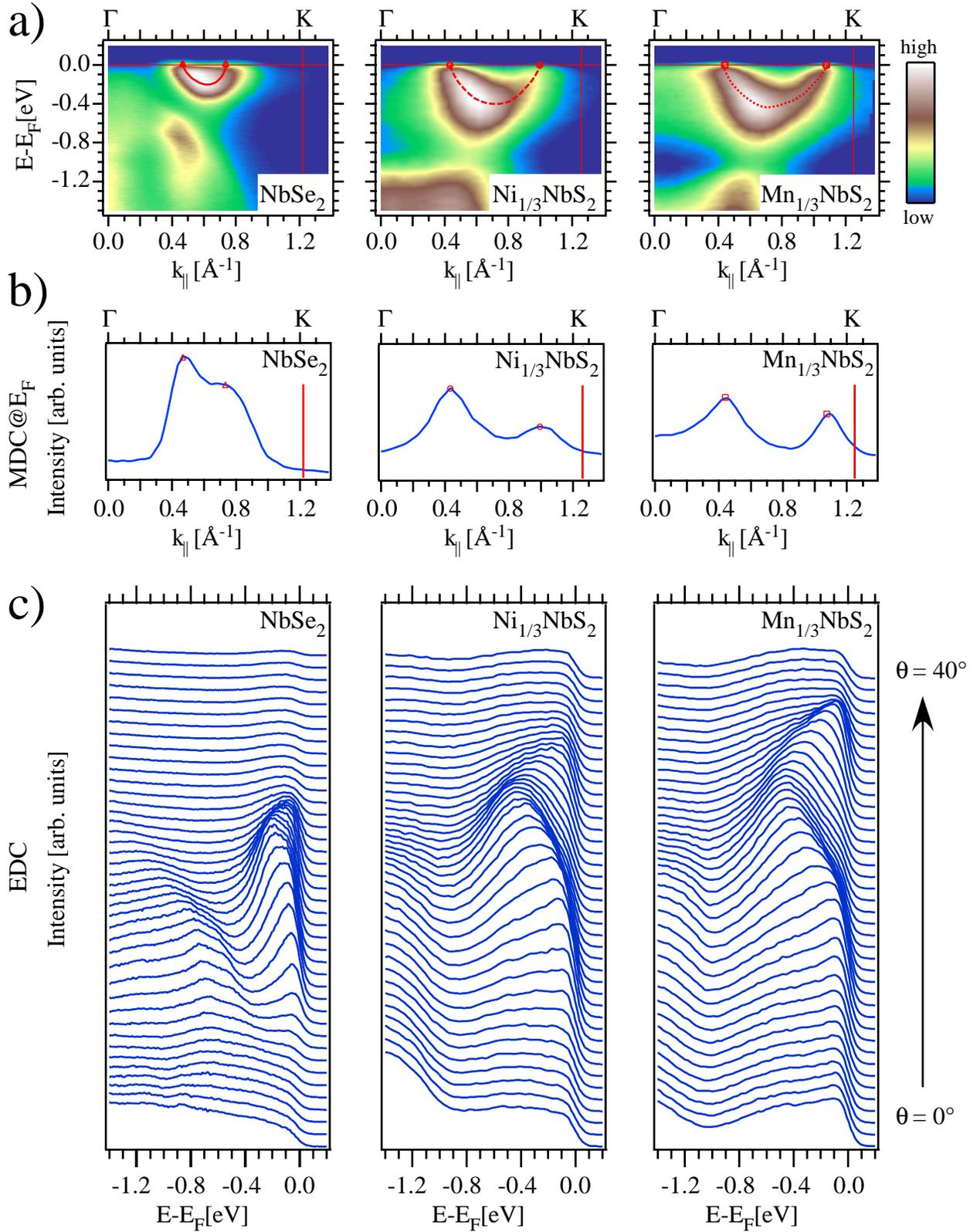}
\caption{a) Comparison between ARPES dispersion maps along the
$\Gamma(A)-K(H)$ direction of NbSe$_2$, Ni$_{1/3}$NbS$_2$ and
Mn$_{1/3}$NbS$_2$ measured at room temperature. b) MDCs extracted at
the Fermi energy $E_F$ as well as c) EDCs stacked as a function of
emission angles $\theta$ are also shown. The red horizontal line
marks the Fermi energy $E_F$, the red vertical line the position of
the $K(H)$ point. To guide the eye, the dispersion of the Nb 4d band
is outlined by red curves. The Fermi vectors obtained from the MDCs
are marked by red symbols. } \label{fig:2}
\end{figure*}

Intercalation of the layered quasi-two-dimensional transition metal
dichalcogenides is possible with a wide variety of electron donor
species ranging from alkali metals \cite{McEwen83} to large organic
molecules \cite{Gamble70}. The intercalation procedure is in general
accompanied by charge transfer from the intercalant species to the
host layer. This allows a fine tuning of the electron occupation of
the relatively narrow d bands defining the Fermi surface of these
compounds. Since the local bonding within the sandwiches is little
changed upon intercalation, the changes in electronic properties are
usually described within the rigid band model, in which the only
change to the host material's electronic structure is the
increased d band filling. \\
The first-row transition metal intercalation complexes are
particularly interesting, because the d electrons left on the
intercalate behave as localized atomic levels with a net magnetic
moment, since there are no adjacent ions to allow overlap and band
formation. Upon cooling, the local moments on these 3d ions
exhibit a variety of magnetic orderings \cite{Parkin80b}.\\
A direct exchange coupling between the moments has been ruled out
because of their large spatial separation \cite{Friend77}. The
anomalous behavior of the Hall coefficient and the resistivity
near the magnetic transition temperature \cite{Parkin80a} suggests
that the conduction electrons play a substantial role in mediating
the exchange interaction between local moments via the
Ruderman-Kittel-Kasuya-Yosida (RKKY) interaction
\cite{Ruderman54,Kasuya56,Yosida57}. In this indirect coupling
mechanism, virtual transitions of the conduction electrons into
the unoccupied orbitals of the 3d ion cause them to experience the
direction of the intercalate moment and result in a local spin
polarization of the conduction electron gas.\\
The response of the conduction electrons to the array of magnetic
moments is determined by the static susceptibility
$\chi(\mathbf{q})$, which depends on the details of the Fermi
surface topology. This is the same susceptibility function which
arises in the Fermi surface nesting criterion for charge density
wave (CDW) formation \cite{Battaglia05}. Any singularity in
$\chi(\mathbf{q})$ at a wavevector $\mathbf{q}$ will give rise to
a spatial oscillation in magnetic polarization of the conduction
electrons away from the 3d ion. Depending on the spin response at
the next 3d ion, the effective coupling may be ferromagnetic or antiferromagnetic.\\
Here, we report on a comparative angle-resolved photoelectron
spectroscopy (ARPES) study of Mn$_{\frac{1}{3}}$NbS$_2$ and
Ni$_{\frac{1}{3}}$NbS$_2$ above and below the magnetic phase
transition temperature. Mn$_{1/3}$NbS$_2$ orders $\;$
ferromagnetically $\;$ at 40 K, while Ni$_{1/3}$NbS$_2$ orders
antiferromagnetically at 90 K. We perform a direct mapping of the
Fermi surface sheets and underlying band structure and compare our
results to the data from 2H-NbSe$_2$. We find that the effect of
intercalation leads to non-uniform doping of the Fermi surface
across the Brillouin zone. We address the impact of our findings on
the nesting behavior of the Fermi surface and discuss the validity
of the rigid band approximation.

\section{Experiment}
\label{sec:1}

ARPES experiments were performed in a modified Vacuum Generator
ESCALAB Mark II spectrometer with a residual gas pressure of
$2\times 10^{-11}$ mbar equipped with a Mg K$_\alpha$ ($\omega=
1253.6$ eV) x-ray anode, a discharge lamp providing
monochromatized He I$\alpha$ ($\omega=21.2$ eV) radiation
\cite{Pillo98}, and a three channeltron hemispherical
electrostatic analyzer kept fixed in space during measurements.
The samples were mounted on a manipulator with two rotational axes
and may be cooled via a closed cycle refrigerator. Energy
resolution is 20 meV, the combined angular resolution of sample
manipulator and analyzer is approximately $1^o$. The different
data acquisition modes are described elsewhere \cite{Clerc06}.\\
Crystals were grown by chemical vapor transport, mounted on the
sample holders using conductive epoxy paste and cleaved in situ
using an aluminum cleaving arm which was fixed onto the sample using
epoxy paste. Despite the fact that the intercalated materials do not
cleave as easily as non-intercalated compounds, we were able in this
way to obtain mirror-like surfaces of sufficient quality. Surface
cleanliness before and after ARPES measurements was monitored by
x-ray photoelectron spectroscopy (XPS). Since the host compound
2H-NbS$_2$ was not available to us, we compare our data to
isostructural and isoelectronic 2H-NbSe$_2$. A tight binding fit
\cite{Doran78} to early non-selfconsistent band structure
calculations \cite{Wexler76} and our own calculations
\cite{Battaglia06} show that the Nb 4d manifold defining the
dominant parts of the Fermi surface is very similar for both
compounds.

\begin{figure}
\includegraphics{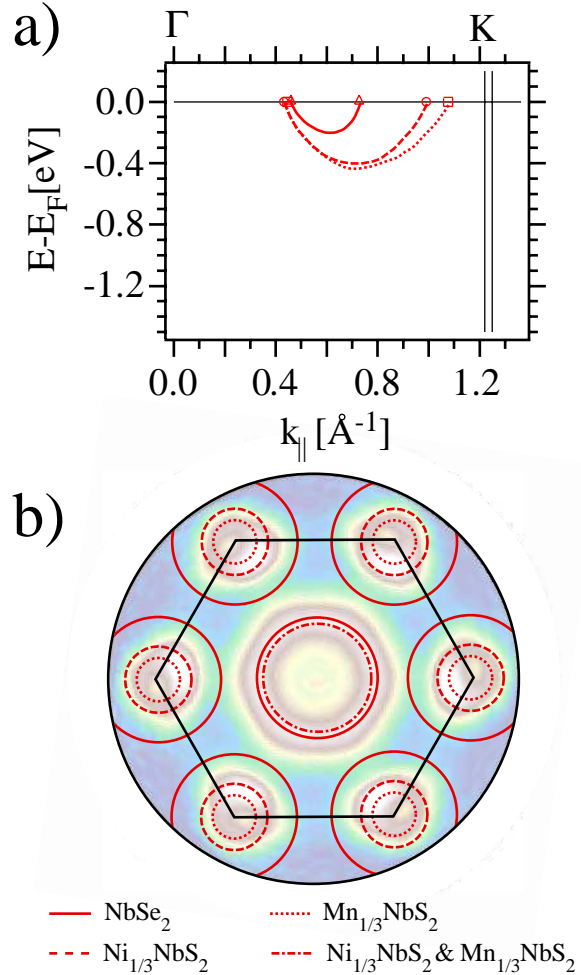}
\caption{Comparison between the Nb 4d band filling of NbSe$_2$,
Mn$_{1/3}$NbS$_2$ and Ni$_{1/3}$NbS$_2$. a) Summary of the
dispersion maps from Fig. \ref{fig:1}a). Note the two positions of
the K point due to the variation in lattice constant during
intercalation. b) Sketch of the effect of doping on the Fermi
surface. } \label{fig:3}
\end{figure}

\section{Results and discussion} \label{sec:2}

Figure \ref{fig:1} shows the intensity distribution of
photoelectrons for Mn$_{1/3}$NbS$_2$ collected from a small,
resolution limited energy window centered on the Fermi energy $E_F$
as a function of the surface-projected electron wave vector
$\mathbf{k}_{||}$. The data has been averaged according to the space
group P$6_322$ \cite{Anzenhofer69} and was divided by a Gaussian
shaped background in order to enhance weaker off-normal emission
features. Apart from a slight variation of the surface-perpendicular
wave-vector $k_\bot$, this map corresponds to a horizontal cut
through the Brillouin zone. It clearly reveals a rounded hexagonal
Fermi surface sheet centered at $\Gamma(A)$ and a second
approximately triangular sheet centered at $K(H)$. Both features are
also observed for NbSe$_2$ \cite{Straub99} and Ni$_{1/3}$NbS$_2$
(not shown). First principles calculations \cite{Corcoran94}
including our own \cite{Battaglia06} show that these features have
predominantly Nb 4d character. For a strictly two-dimensional solid,
the electronic dispersion is completely determined by
$\mathbf{k}_{||}$, because there is no dispersion along $k_\bot$.
Recent ARPES measurements have explored the $k_\bot$ dependence of
the Fermi surface of NbSe$_2$ by varying the excitation energy and
revealed a high degree of two dimensionality of the Nb 4d Fermi
surface cylinders \cite{Rossnagel01}. Because the unit cell of the
host compound 2H-NbS$_2$ contains two formula units, all bands and
hence the Fermi surface sheets are actually doubled. Their
degeneracy is lifted by interlayer coupling and spin-orbit
interaction. Although the double-walled nature of the two Fermi
surface sheets is not directly observed in the Fermi surface maps,
it was shown earlier for NbSe$_2$ that the conduction band doublet
is in fact resolvable by ARPES in energy distribution curves (EDC)
\cite{Straub99}. However a full discussion of the lineshape of the
spectral function of this class of materials is not subject of the
present
study and can be found in Ref. \cite{Clerc06,Clerc07,Battaglia05}.\\
In Fig. \ref{fig:2}a) we compare ARPES dispersion maps measured
along $\Gamma(A)-K(H)$ of Mn$_{1/3}$NbS$_2$ and Ni$_{1/3}$NbS$_2$
with data from NbSe$_2$. Except for a sharpening of the Fermi edge,
the spectra acquired below 20 K show the same behavior as the room
temperature data within our angular and energy resolution. The color
scale represents the intensity of the emitted photoelectrons plotted
as a function of energy $E$ and crystal momentum $\mathbf{k}_{||}$.
Since the in-plane momentum of the photoelectron is conserved during
the photoemission process, $\mathbf{k}_{||}$ of the electrons in the
solid is obtained from the emission angle $\theta$ via
$\mathbf{k}_{||}=\sqrt{2m (\omega+E-E_F-\phi_s)} sin{\theta}$ where
$m$ is the electron mass, $\omega$ the excitation energy and
$\phi_s$ the sample work function \cite{workfunction}. Corresponding
EDCs are shown in Fig. \ref{fig:2}c). As a guide to the eye the
dispersion of the Nb 4d band is outlined by red curves and
summarized in Fig. \ref{fig:3}a). Intensity at higher binding energy
originates from S/Se 4p derived states. The Nb 4d band possesses an
approximately parabolic dispersion for all three compounds. As
expected within the rigid band picture, the main effect of the
intercalation is the increased band filling reflected by a shift of
the band bottoms of approximately 200 meV towards higher binding
energies with respect to NbSe$_2$ for both intercalated compound.
Simultaneously the dispersion parabola shifts away from the
$\Gamma(A)$ point ($\theta=0^o$) towards the $K(H)$ point
($\theta=35^o$ for NbSe$_2$ and $36^o$ for the intercalated
compounds) violating the rigid band approximation. The locations of
the Fermi crossings $\mathbf{k}_F$ were obtained by fitting two
Lorentzian shaped peaks to the momentum distribution curves (MDC)
extracted from the dispersion maps at the Fermi energy shown in Fig.
\ref{fig:2}b). It is interesting to note that the Fermi point at
$\mathbf{k}_{||}=0.44-0.46$ \AA$^{-1}$, defining the size of the
hexagonal hole pocket centered around the $\Gamma(A)$ point, is not
affected by intercalation within our experimental resolution (our
angular resolution of $\Delta\theta=1^o$ translates into an
uncertainty of $\Delta\mathbf{k}_{||}\leq 0.04$ \AA$^{-1}$), whereas
the second Fermi point strongly shifts towards the Brillouin zone
border, causing the triangular pockets around $K(H)$ to shrink
considerably. Thus intercalation leads to non-uniform, pocket
selective doping of the Fermi surface, reducing the size of the
$K(H)$ centered hole pockets, but leaving the occupation of the
$\Gamma(A)$ pocket approximately unchanged. A sketch of this
situation is shown in Fig. \ref{fig:3}b). At
present we are not able to explain why only the $K(H)$ pockets are doped.\\
Assuming a strictly two-dimensional dispersion and cylindrical hole
pockets as shown in Fig. \ref{fig:3}b), we estimate the electron
filling of the Nb 4d band by computing the ratio between the area of
occupied states and the area of the entire Brillouin zone. For the
non-intercalated NbSe$_2$ compound we obtain a filling of 63 $\%$.
Due to the stochiometry and neglecting the bilayer splitting
discussed before, therefore considering only one single band, the
filling should be precisely 50 $\%$. Our value is however reasonable
since the existence of an additional small pancake shaped, Se
derived, hole sheet centered around the $\Gamma(A)$ point is
predicted by theory and has been confirmed by experiment
\cite{Corcoran94,Straub99,Rossnagel01}. The measured filling
increases upon intercalation to 79 $\%$ and 82 $\%$ for the Ni and
Mn intercalated compound respectively. In a simple ionic picture,
assuming that Ni and Mn are divalent, the resulting d band filling
of 5/6=83 $\%$ slightly overestimates the charge transfer from the
intercalant to the Nb d band \cite{filling}. The valence state of
the Ni and Mn ion determined from the band filling is +1.74 and
+1.92 respectively. This behavior is consistent with the higher
ionization energy of Ni with respect to Mn \cite{Page90,Sugar85}.
The valence state of +2.01 for the Mn ion derived from an earlier
ARPES study \cite{Barry83b,Barry83} is also in good agreement with
our result. Furthermore the multiplet splitting of the intercalant
3s core levels probed by x-ray photoemission spectra \cite{Barry83b}
is consistent with the adoption of an approximate +2 valence state
of the magnetic ions. Numerical values concerning the evolution of
the Nb 4d band upon intercalation are summarized in table
\ref{tab:1}.
\\
\begin{table}
\caption{Evolution of the Nb 4d band upon intercalation}
\label{tab:1}       
\begin{tabular}{llll}
\hline\noalign{\smallskip}
compound& bottom of band & Fermi crossing $\mathbf{k}_F$& band filling  \\
&[eV]& [\AA$^{-1}$]&[$\%$] \\
\noalign{\smallskip}\hline\noalign{\smallskip}
NbSe$_2$ & -0.2 eV & 0.46, 0.74 & 0.63 \\
Ni$_{1/3}$NbS$_2$ & -0.4 eV & 0.44, 1.00 & 0.79 \\
Mn$_{1/3}$NbS$_2$ & -0.4 eV & 0.44, 1.08 & 0.82 \\
\noalign{\smallskip}\hline
\end{tabular}
\end{table}
We now address the issue of the validity of the doping description
traditionally adopted for the interpretation of these compounds.
Since in Mn$_{1/3}$NbS$_2$ and Ni$_{1/3}$NbS$_2$, the intercalant
species occupy well defined interlayer sites forming a hexagonal
$\sqrt{3}\times\sqrt{3}$ superlattice, one could alternatively
interpret the intercalated compounds as stochiometric materials
containing three formula units per unit cell. The superlattice gives
rise to an additional periodic potential, which, within the Bloch
theory of periodic crystals, is expected to fold back dispersion
branches into the corresponding smaller Brillouin zone shown in Fig.
\ref{fig:1}. The relevance of this reduced zone scheme however
remains unclear for a variety of compounds
\cite{Battaglia05,Voit00}. For the compounds under investigation, we
do not find clear evidence for backfolding consistent with an
earlier ARPES study \cite{Barry83} indicating that the superlattice
potential is weak. We also note that the Nb 4d band minimum of the
intercalated compounds does not fall on the $M$ point of the new
Brillouin zone, called $M'$ in Fig. \ref{fig:1}a), which is located
halfway between the $\Gamma$ and $K$ point of the large Brillouin
zone, violating the strict requirement of the Bloch theory that
electron bands must cross the Brillouin zone with zero velocity.
These observations thus support a description of the intercalation
process via doping of the parent band structure.
\\
We now turn the discussion to the implications of the observed
pocket selective doping on the nesting properties of the Fermi
surface which via the susceptibility function determine the strength
and range of the RKKY interaction. The doping independent
$\Gamma(A)$ centered hexagonal hole cylinder provides a favorable
topology for a threefold degenerate nesting vector directed along
the $\Gamma-M$ direction corresponding approximately to a $(3 \times
3)$ superlattice in real space (see Fig. \ref{fig:1}b).  A $(3
\times 3)$ magnetic superlattice is commensurate with the
$(\sqrt{3}\times\sqrt{3})$ superlattice formed by the intercalate
moments and thus compatible with a magnetically ordered state. While
the RKKY interaction decays isotropically as $R^{-3}$ in the free
electron case, $\mathbf{R}$ being the vector between two magnetic
ions, the interaction becomes longer ranged for an anisotropic Fermi
surface topology \cite{Roth66}. For the special case of a
cylindrical region of the Fermi surface, the decay rate is governed
by $R^{-2}$ along the direction perpendicular to the axis of the
cylinder. For two flat parallel regions of the Fermi surface, i.e.
for ideal nesting conditions, the interaction is found to fall off
only as $R^{-1}$ in the direction perpendicular to the two planes.
Furthermore, the sign of the RKKY coupling is modulated by
$\sin{(k_{z}-k_{z}')R}$, where $\mathbf{k}$ and $\mathbf{k'}$ are
two points on the Fermi surface and $z$ is chosen along the $R$
direction, allowing either antiferromagnetic or ferromagnetic
coupling depending on the Fermi surface topology. However, if the
RKKY interaction were the only important interaction and the
relevant nesting took place within the $\Gamma(A)$ centered pocket,
the magnetic behavior would be the same for both the Ni and the Mn
intercalated compound. This is inconsistent with the experimental
results since the Ni intercalates order antiferromagnetically, the
Mn intercalates ferromagnetically. The nesting vector for the
$\Gamma(A)$ centered sheet approximately coincides with the CDW
vector observed by neutron scattering for NbSe$_2$ \cite{Moncton75}.
But because no evidence for a CDW-induced gap opening in ARPES
spectra of NbSe$_2$ was found \cite{Straub99,Rossnagel01}, the
driving mechanism for the CDW transition has remained controversial
and the simple nesting scenario has been questioned. It is
instructive that in 2H-TaSe$_2$, which exhibits a similar Fermi
surface topology, the CDW induced energy gap is nearly zero for the
$\Gamma(A)$ centered sheet, although this sheet exhibits comparable
nesting qualities as for NbSe$_2$. Instead the opening of a gap is
observed on the $K(H)$ centered pocket \cite{Liu00}. This
observation led to the suggestion \cite{Valla04} that the CDW state
originates from the $K(H)$ centered sheet. Nesting across these
pockets would possibly explain the occurrence of different magnetic
orders, since the nesting vector depends on the band filling, which
in turn depends on the intercalant species. However, since the
cross-section of these cylinders are rounded triangles with flat
edges oriented at 120$^o$ with respect to one another (see Fig.
\ref{fig:1}a), substantial nesting is highly unlikely. According to
a recent first principle study for NbSe$_2$ \cite{Johannes06} strong
nesting occurs only between the flat triangular edges of the $K(H)$
pockets and the parallel flat edges of the central hexagonal pocket,
resulting in a nesting vector along the $\Gamma-K$ direction. This
nesting vector would be doping dependent and could explain the
variations in magnetic order. Based on the observation that the
anomalies in the transport data for Ni$_{1/3}$NbS$_2$ are less
pronounced than for Mn$_{1/3}$NbS$_2$, which indicates a weaker
coupling of the conduction electrons to the magnetic moments, an
alternative explanation has been considered: an additional
superexchange interaction via the orbitals of the non-magnetic
sulfur atoms is expected to lead to a predominantly
antiferromagnetic coupling \cite{Parkin80a}. In this framework it is
considered that the superexchange interaction is small for Mn
intercalate, but becomes progressively larger as the intercalate is
varied from Mn to Ni. The interplay between RKKY and superexchange
interaction and the relevance of a nesting scenario should be
revised taking into account our experimental observation of the
doping dependence of the $K(H)$ pockets. For a final picture
additional experimental data is required.

\section{Conclusion} \label{concl}

We have performed full-hemispherical Fermi surface mapping of
Mn$_{1/3}$NbS$_2$ and Ni$_{1/3}$NbS$_2$ and validate the doping
description of the intercalation process. Two hole pockets, one
centered at the $\Gamma(A)$ point, the second centered at the $K(H)$
point are observed. Doping due to the intercalation of the host
compound NbS$_2$ is non-uniform across the Brillouin zone and causes
the $K(H)$ pockets to shrink, while the filling of the zone centered
hole pockets remains surprisingly unchanged. Thus the rigid band
model should be applied carefully and may only serve as a first
approximation. The doping dependence of the nesting vector could
possibly explain the different magnetic orders observed for the Mn
and Ni compounds, but a simple nesting scenario does not appear to
be sufficient for a complete picture and further complementary
investigations are required.

\section{Acknowlegdments}

The help of Leslie-Anne Fendt, Hans Beck, Samuel Hoffmann and
Christian Koitzsch is gratefully acknowledged. Skillfull technical
assistance was provided by our workshop and electric engineering
team. This work was supported by the Fonds National Suisse pour la
Recherche Scientifique through Div. II and MaNEP.

%
%
%
%
 \bibliographystyle{unsrt}
 \bibliography{NiNb3S6}

\begin{thebibliography}{10}
\bibitem{workfunction}
Due to a possible modification of the sample workfunction during the
  intercalation process, $\mathbf{k}_{||}$ slightly varies. However, for a
  typical value of the workfunction $\phi=4$ eV with an associated error of
  $\Delta\phi=1$ eV, variations of $k_{||}$ are below 3$\%$ and do not
  influence our conclusions.
\bibitem{filling}
In the purely ionic picture, the two chalcogen atoms per formula unit can fill
  their outer p shell by taking four electrons provided by the Nb atom. This
  leads to a d$^1$ configuration with a single electron left in the Nb d shell.
  Assuming a divalent state for the intercalant ion and taking into account
  that only $\frac{1}{3}$ of the ion is accounted for per formula unit, it
  contributes an additional $2\times\frac{1}{3}$ electrons per formula unit to
  this band, leading to a total band filling of
  $\frac{1}{2}(1+\frac{2}{3})=\frac{5}{6}$, where the factor $\frac{1}{2}$
  takes into account the fact that each band can accommodate two electrons.
\end{thebibliography}
%
%
%

\end{document}